\documentclass[letter]{aa}
\usepackage{graphicx}
\usepackage{txfonts}
\usepackage{multirow}
\usepackage{media9}
\usepackage{animate}

\begin{document}

   \title{The Venus' Cloud Discontinuity in 2022}

   \subtitle{A first long-term study with uninterrupted observations}

   \author{J. Peralta\inst{1}\fnmsep\thanks{Corresponding author} \and
           A. Cidad\~{a}o\inst{2} \and
           L. Morrone\inst{3}\fnmsep\inst{4} \and
           C. Foster\inst{5} \and
           M. Bullock\inst{6} \and
           E. F. Young\inst{7} \and
           I. Garate-Lopez\inst{8} \and
           A. S\'{a}nchez-Lavega\inst{8} \and
           T. Horinouchi\inst{9} \and
           T. Imamura\inst{10} \and
           E. Kardasis\inst{11} \and
           A. Yamazaki\inst{12} \and
           S. Watanabe\inst{13}
          }

   \institute{Facultad de Física, Universidad de Sevilla, Av. Reina Mercedes s/n, 41012 Sevilla, Spain\\
              \email{jperalta1@us.es} \and
             Associaç\~{a}o Portuguesa de Astr\'{o}nomos Amadores (APAA), Portugal \and
             AstroCampania, Italy \and
             British Astronomical Association, UK \and
             Astronomical Society of Southern Africa (ASSA), South Africa \and
             Science and Technology Corp., Boulder, Colorado, USA \and
             Southwest Research Institute, Boulder, Colorado, USA \and
             Escuela de Ingenier\'{i}a de Bilbao, Universidad del Pa\'{i}s Vasco (UPV/EHU), Bilbao, Spain \and
             Faculty of Environmental Earth Science, Hokkaido University, Sapporo, Japan \and
             Graduate School of Frontier Sciences, The University of Tokyo, Japan \and
             Hellenic Amateur Astronomy Association, Athens, Greece \and
             Institute of Space and Astronautical Science, Japan Aerospace Exploration Agency (JAXA), Sagamihara, Japan \and
             Hokkaido Information University, Ebetsu, Japan
             }

   \date{Received August 29, 2022; accepted February 07, 2023}

 
  \abstract
   {First identified in 2016 by the Japan Aerospace eXploration Agency (JAXA) Akatsuki mission, the discontinuity/disruption is a recurrent wave observed to propagate during decades at the deeper clouds of Venus (47--56 km above the surface), while its absence at the clouds' top ($\sim$70 km) suggests that it dissipates at the upper clouds and contributes in the maintenance of the puzzling atmospheric superrotation of Venus through wave-mean flow interaction.}
   {Taking advantage of the campaign of ground-based observations undertaken in coordination with the Akatsuki mission since December 2021 until July 2022, we aimed to undertake the longest uninterrupted monitoring of the cloud discontinuity up to date to obtain a pioneering long-term characterization of its main properties and better constrain its recurrence and lifetime.}
   {The dayside upper, middle and nightside lower clouds were studied with images with suitable filters acquired by Akatsuki's Ultraviolet Imager (UVI), amateur observers and SpeX at NASA's Infrared Telescope Facility (IRTF). Hundreds of images were inspected in search of the discontinuity events and to measure key properties like its dimensions, orientation or rotation period.}
   {We succeeded in tracking the discontinuity at the middle clouds during 109 days without interruption. The discontinuity exhibited properties nearly identical to measurements in 2016 and 2020, with an orientation of $91^{\circ}\pm 8^{\circ}$, length/width of $4100\pm 800$ / $500\pm 100$ km and a rotation period of $5.11\pm 0.09$ days. Ultraviolet images during 13-14 June 2022 suggest that the discontinuity may have manifested at the top of the clouds during $\sim$21 hours as a result of an altitude change in the critical level for this wave due to slower zonal winds.}
   {}


   \maketitle
%

\section{Introduction}
From the surface up to $\sim$90 km. the general circulation of the Venus's atmosphere is dominated by a retrograde zonal superrotation whose speeds peak at the region of the clouds (45--70 km). Nevertheless, key aspects about its generation and maintenance are yet poorly understood \citep{Sanchez-Lavega2017} and present general circulation models (h.a. CGMs) yet fail to accurately reproduce it \citep{Lebonnois2013,Navarro2021}. Recent works have provided new and solid evidences of the importance of planetary-scale waves at the upper clouds like the so-called Y-feature \citep{Boyer1961,Kouyama2012,Peralta2015} or the solar tides excited at the level in maintaining the superrotation \citep{Kouyama2015,Kouyama2019,Horinouchi2020,Fukuya2021}. Besides, the waves excited below the clouds have been in the spotlight during the last years too, thanks to recent reports about new waves discovered by the ongoing JAXA’s Akatsuki mission \citep{Nakamura2016} and earlier ESA's Venus Express \citep{Svedhem2007}. Numerous stationary waves excite at the surface have been reported in images of the thermal emission of the upper clouds \citep{Fukuhara2017NatGeo,Peralta2017NatAstro}, 283-nm albedo at the clouds' top \citep{Kitahara2019} or near infrared images at 2.02 $\mu$m \citep{Sato2020}.\\
\\
Images of Venus sensing the middle and the lower clouds have revealed a disruption or discontinuity on their albedo/opacity \citep{Peralta2020,Kardasis2022} linked to dramatic changes on the clouds' optical thickness and distribution of aerosols \citep{McGouldrick2021}, and shown to be a recurrent atmospheric phenomenon missed for decades. Based on simulations with the Institut Pierre Simon Laplace (IPSL) Venus GCM \citep{Scarica2019}, it has been suggested that the discontinuity may be a new type of Kelvin wave feeding the superrotation with momentum from the deeper atmosphere, while its phase speeds along with its absence in observations of the upper clouds evidenced that this Kelvin wave might find its critical level below the top of the clouds \citep{Peralta2020,Kardasis2022}. But despite the potential relevance for this new wave, we have not been able to determine key aspects like its source of excitation, the excitation/dissipation heights, the long-term behaviour and its recurrence in the Venus atmosphere.\\
\\
Taking advantage of an excellent campaign of Venus observations undertaken during the year 2022 with a combined effort from JAXA's Akatsuki mission \citep{Nakamura2007,Nakamura2016}, NASA's Infrared Telescope Facility IRTF \citep{Rayner2003} and amateur observers, we present the results for a long-term continuous monitoring of the clouds' discontinuity and the first report of its propagation up to the top of the clouds. A description of the data sets and methods is introduced in section \ref{methods}, while the results and the discussion about the measured properties of the discontinuity and its manifestation at the top of the clouds can be consulted in section \ref{results}. We finish with the main conclusions in section \ref{conclusions}.

\section{Data sets and Methods}\label{methods}
As in previous works \citep{Peralta2020,Kardasis2022}, we searched for events of the discontinuity in images of Venus sensing the three main layers of the clouds \citep{Titov2018}: the dayside upper clouds (56.5–70 km above the surface) whose albedo at $\sim$70 km can be observed with images taken at ultraviolet wavelengths \citep{Ignatiev2009,Limaye2018EPS}, the dayside middle clouds (50.5–56.5 km) which can be observed at visible and near-infrared wavelengths \citep{Hueso2015,Peralta2019GRL}, and the nightside lower clouds (47.5–50.5 km) whose transparency/opacity to the deeper atmospheric thermal emission is observable in infrared atmospheric windows at 1.74, 2.26 and 2.32 $\mu$m \citep{Satoh2017,Peralta2018ApJS}.

\subsection{Images obtained with IRTF/SpeX}
The nightside lower clouds of Venus were explored using images of Venus taken from NASA's IRTF (Mauna Kea, Hawaii) with the guide camera of the instrument SpeX \citep{Rayner2003} and contK (2.26 $\mathrm{\mu m}$) and 1.74-$\mathrm{\mu m}$ filters. Our observations\footnote{Data available upon reasonable request.} covered the following dates: 2021 December 3--19 (12 nights), 2022 January 27--February 23 (23 nights), and 2022 March 5--12 (6 nights). During this period, the phase angle of Venus varied within 140$^{\circ}$--96$^{\circ}$ and the apparent size ranged 52--27 arcsec. Since the plate scale of SpeX is 0.116 arcsec per pixel, the spatial resolution of Venus in these images varied from 42 to 82 km per pixel. As in previous works \citep{Peralta2019Icarus,Peralta2020}, IRTF/SpeX images were corrected for the saturated dayside, stacked, navigated and processed following the same procedure originally described in detail by \citet{Peralta2018ApJS}. Unfortunately, an accurate radiometric calibration was not possible due to a persistent problem of light contamination from the saturated dayside of the planet similar to the one affecting images by Akatsuki/IR2 \citep{Satoh2022}.

\subsection{Images obtained by amateur observers}
To study the presence of the discontinuity on the dayside middle clouds, we used images of Venus acquired by amateur observers located in Italy, Portugal, South Africa, Australia, Estonia and Taiwan. We obtained the images directly from some of the observers or through the Venus section of ALPO-Japan\footnote{http://alpo-j.sakura.ne.jp/Latest/Venus.htm} \citep{Sato2018}. These consist of stacks of thousands of images acquired during time intervals rarely longer than 5 minutes \citep[see lucky imaging procedure in][]{Kardasis2022}, and they were captured using small telescopes with diameters typically ranging 203--355 mm, observing at wavelengths $>685$ nm in all the cases by means of high-pass and band-pass filters \citep[most used filters are detailed in Table 1 of][]{Kardasis2022}. After a careful selection, we inspected a total of 319 images covering nearly every day since 2022 March 13 until July 31. During this period, the phase angle of Venus varied within 95$^{\circ}$--32$^{\circ}$ and the apparent size ranged 27--11 arcsec. For wavelengths of $\sim$700 nm and telescope of 355-mm diameter we would expect a nominal resolution of $\sim$0.5 arcsec for each individual image, and 350--860 km per pixel along the mentioned observing period. Since the lucky imaging has been demonstrated superresolution capabilities and improve the resolution up to a factor of four in poor seeing conditions \citep{Farsiu2004,Law2006,Sanchez-Lavega2016,Peralta2017NatAstro}, we can assume a conservative improvement of a factor of two for the stacked images, implying spatial resolutions within 175--430 km per pixel.

\subsection{Images obtained with Akatsuki/UVI and LIR}
The level of the upper clouds was studied with images taken by two instruments onboard the Akatsuki orbiter: 365 and 283-nm images by the Ultraviolet Imager (UVI) \citep{Yamazaki2018} were used to explore the cloud patterns and zonal winds at the top of the dayside upper clouds ($\sim$70 km) while 10-$\mathrm{\mu m}$ images by the Longwave Infrared camera (LIR) can sense thermal contrasts within a wider altitude range of 60--75 km \citep{Taguchi2007} which vary with observed emission angle \citep{Kouyama2017,Akiba2021}. UVI and LIR data consisted of L2b and L3bx products \citep{Ogohara2017,Murakami2019} from the mission internal release\footnote{Eventually available via Data ARchives and Transmission System (DARTS) by ISAS/JAXA Center for Science-satellite Operation and Data Archive (C-SODA), and NASA PDS Atmospheres Node.} v20220901, and the images matched the same observing period as for the middle clouds with amateur observations. The imagery data set finally selected covered from 2022 April 15 until July 30 (Akatsuki orbits 212--221), with the spatial resolution at lower latitudes of Venus varying 75--10 km per pixel in UVI images and 320--40 km per pixel in LIR images. We explored a total of 928 365-nm UVI images (similar number of 283-nm images) and 623 LIR images, although in the latter case we only examined LIR images acquired during events of the discontinuity (see subsection \ref{properties}).

\begin{figure*}[t]
\centering
   \includegraphics[width=18cm]{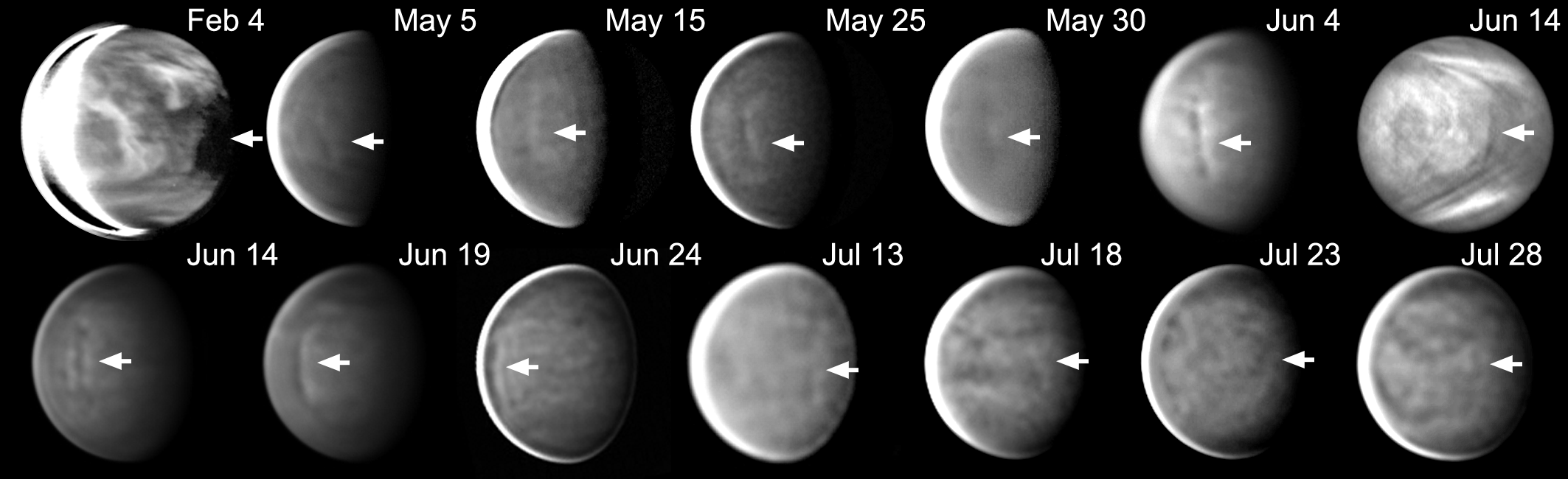}
     \caption{Examples of discontinuity events during 2022. The discontinuity was apparent on the nightside lower clouds with $2.26-\mathrm{\mu m}$ images from IRTF/SpeX taken in February 4, and at the dayside middle clouds as observed on images by amateur observers taken with filters covering wavelengths 700--900 nm from May to July 2022. During June 14, Akatsuki/UVI images at 365 nm suggest that the discontinuity was propagating simultaneously at the upper and middle clouds (last image in upper row). All the images were high-pass filtered to enhance cloud details.}
     \label{fig:discontinuities}
\end{figure*}

\subsection{Measurement of speeds}
Prior to measurements, all the images were geometrically projected onto an equirectangular (cylindrical) geometry with an angular resolution equivalent to the best resolution in the original images. To measure winds and the phase speed of the discontinuity, we applied the manual method by \citet{Sanchez-Lavega2016} and another technique consisting on a manual search of cloud tracers followed by a fine adjustment using automatic template matching based on phase correlation, which is visually accepted or rejected by a human operator \citep[subsection 2.3]{Peralta2018ApJS}.\\
\\
Towards a more accurate characterization of the difference between the discontinuity phase speed and zonal winds \citep{Peralta2020,Kardasis2022}, the phase speed of the discontinuity was measured by comparing its position after one or two full revolutions around the Venus globe, allowing to achieve an accuracy of 0.2--2.7 m$\cdot$s$^{-1}$. Concerning the measurement of wind speeds at the nightside lower clouds, we combined pairs of stacked images from SpeX typically separated by 90--155 minutes which enabled us to infer winds with errors within 3--8 m$\cdot$s$^{-1}$. In the case of the UVI 365-nm images, we obtained wind speeds with errors ranging 2--11 m$\cdot$s$^{-1}$ by considering pairs of images acquired when Akatsuki approached/receded its orbital pericenter or when cloud tracers could be tracked during 3--5 hours. We excluded our wind measurements for dayside middle clouds since the worse spatial resolution of amateur images generally precluded speeds with errors below 40 m$\cdot$s$^{-1}$. The speeds in section \ref{results} were calculated as the mean and standard deviation for speeds of individual tracers between 30$^{\circ}$N--30$^{\circ}$S.

\section{Results}\label{results}
\subsection{Properties of the discontinuity}\label{properties}
\begin{figure}
\centering
   \includegraphics[width=\columnwidth]{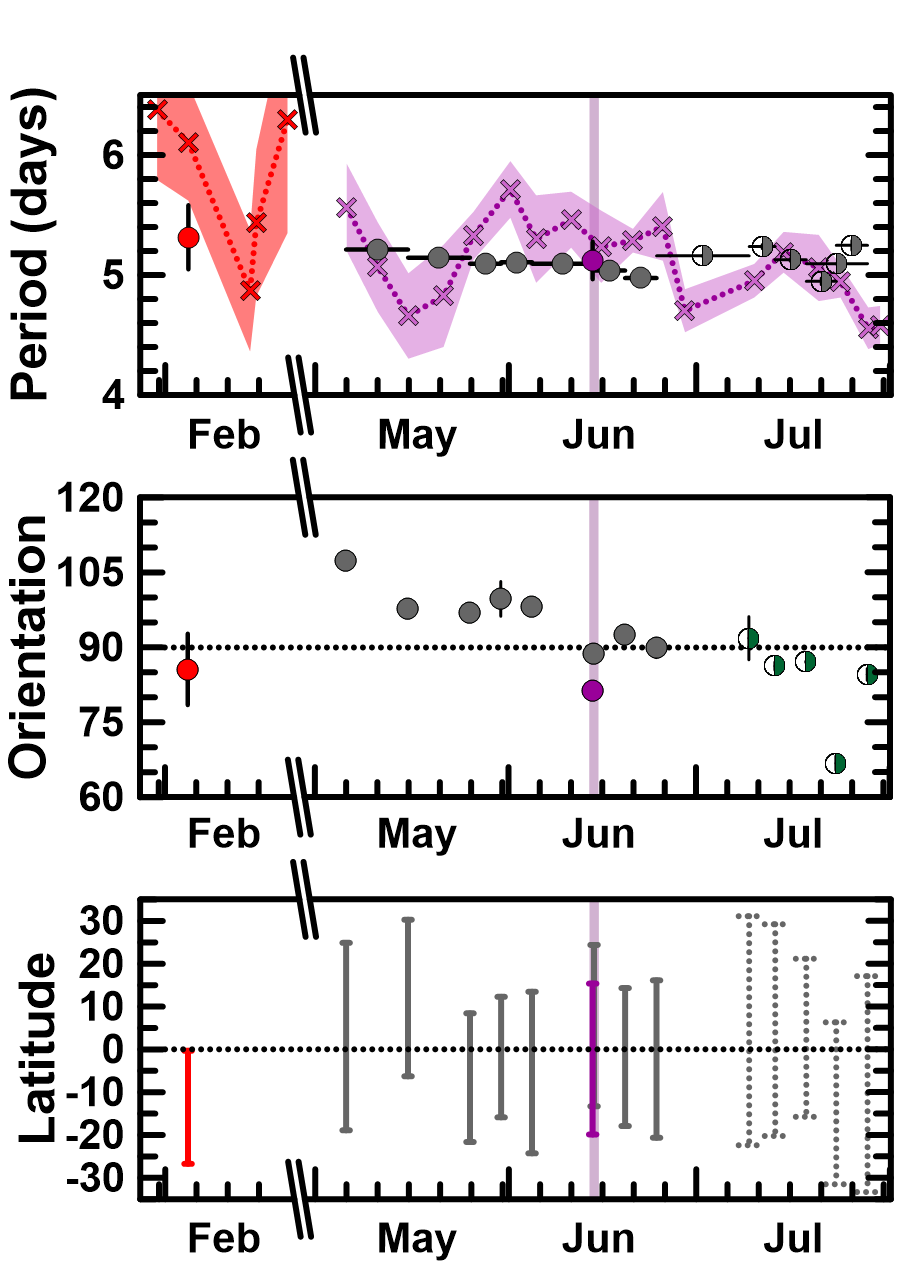}
     \caption{Properties of the discontinuity during 2022. Panel \textbf{(a)} exhibits the rotation period (as terrestrial days) for the discontinuity zonal phase speed (marked with circles) and zonal winds (crosses) along with their errors (bars for phase speeds, shadowed areas for winds). Red colour represents winds and discontinuity events at the nightside lower clouds, grey and purple correspond to dayside middle and upper clouds, respectively. Panel \textbf{(b)} displays the orientation of the discontinuity (circles) as degrees relative to equator/parallels. Panel \textbf{(c)} shows the meridional coverage of the discontinuity (vertical bars). Since the discontinuity seemed to weaken during July 2022, data is displayed as half-filled circles (period and orientation) and dotted bars (meridional coverage). The time frame in light violet corresponds to the event when the discontinuity simultaneously manifested at the upper and middle clouds.}
     \label{fig:CD-properties}
\end{figure}

We observed the nightside lower clouds of Venus between 2021 December 3 and 2022 March 12, while the dayside clouds were studied between 2022 March 13 and July 31. Table \ref{table:events} describes the discontinuity events identified during these campaigns, while Figs.~\ref{fig:discontinuities} and \ref{fig:s-discontinuities} exhibit images of these events. Despite the number of observations with IRTF/SpeX between December 2021 until March 2022, only one positive event was confirmed with the clouds' discontinuity with a length/width of $2900\pm 700$ / $400\pm 40$ km, an orientation of $91^{\circ}\pm 8^{\circ}$, and propagating with an phase speed of $-79\pm 6$ m$\cdot$s$^{-1}$ ($10\pm 8$ m$\cdot$s$^{-1}$ faster than winds), and faster than the speed $-68\pm 9$ m$\cdot$s$^{-1}$ in 2016--2018 \citep{Peralta2020}. An abrupt change in the clouds' opacity was also observed in February 15 and 20 (see Fig.\ref{fig:s-CD-LowerClouds}), suggesting the reappearance of the first discontinuity after several full revolutions. Nevertheless, their zonal speeds matched those of the background winds within error bars and we decided to discard them from the analysis. The fast increase of zonal wind speeds after February 4 (see Fig.~\ref{fig:CD-properties}a) suggests a probable dissipation of the wave leaving a persistent reminiscence of its effect on the clouds' opacity.\\

\begin{table}
\caption{Discontinuity events observed during February--July 2022}     
\label{table:events}      
\centering                                      
\begin{tabular}{c c c c}          
\hline\hline                        
Date/Time (UT) & Clouds\tablefootmark{a} & Wavelength\tablefootmark{b} & Observer\\    
\hline                                   
    2022-02-04 16:55 & Lower & 2.26 $\mathrm{\mu m}$ & IRTF/SpeX \\      
    2022-05-05 04:36 & Middle & $>$742 nm & C. Foster    \\
    \phantom{2022-05-05} 06:13 & Middle & 820--920 nm & A. Cidad\~{a}o    \\
    2022-05-15 04:29 & Middle & $>$742 nm & C. Foster    \\
    \phantom{2022-05-15} 08:16 & Middle & 820--920 nm & A. Cidad\~{a}o    \\
    2022-05-25 04:41 & Middle & $>$742 nm & C. Foster    \\
    \phantom{2022-05-25} 07:58 & Middle & 820--920 nm & A. Cidad\~{a}o    \\
    2022-05-30 05:14 & Middle & $>$742 nm & C. Foster    \\
    \phantom{2022-05-30} 09:06 & Middle & $>$685 nm & L.~S. Viola    \\
    2022-06-04 05:09 & Middle & 820--920 nm & L. Morrone    \\
    \phantom{2022-06-04} 09:52 & Middle & 820--920 nm & A. Cidad\~{a}o    \\
    2022-06-13 14:05 & \multirow{2}{*}{Upper} & \multirow{2}{*}{345--380 nm} & \multirow{2}{*}{VCO\tablefootmark{c}/UVI}    \\
    2022-06-14 23:05 &  &  &     \\
    2022-06-14 05:16 & Middle & 820--920 nm & L. Morrone    \\
    2022-06-19 05:52 & Middle & 820--920 nm & L. Morrone    \\
    \phantom{2022-06-19} 06:48 & Middle & 820--920 nm & A. Cidad\~{a}o    \\
    2022-06-24 05:46 & Middle & 820--920 nm & L. Morrone    \\
    \phantom{2022-06-24} 07:36 & Middle & 820--920 nm & A. Cidad\~{a}o    \\
    2022-07-09 06:33 & Middle & 820--920 nm & A. Cidad\~{a}o    \\
    2022-07-13 10:33 & Middle & 820--920 nm & A. Cidad\~{a}o    \\
    2022-07-18 11:57 & Middle & 820--920 nm & A. Cidad\~{a}o    \\
    2022-07-23 06:27 & Middle & 820--920 nm & L. Morrone    \\
    \phantom{2022-07-23} 07:54 & Middle & 820--920 nm & A. Cidad\~{a}o    \\
    2022-07-28 09:21 & Middle & 820--920 nm & A. Cidad\~{a}o    \\
\hline                                             
\end{tabular}
\tablefoot{
\tablefoottext{a}{Approximate level sensed: the dayside upper clouds (56.5–70 km above the surface), the dayside middle clouds (50.5–56.5 km), and the nightside lower clouds (47.5–50.5 km).}\\
\tablefoottext{b}{Filters used: contK (2.22--2.32 $\mathrm{\mu m}$), Astronomik ProPlanet 742 IR-pass filter ($>$742 nm), Baader SLOAN/SDSS z'-s' (820--920 nm), Baader IR-Pass Filter ($>$685 nm) and VCO/UVI 365-nm (345--380 nm).}\\
\tablefoottext{c}{Venus Climate Orbiter: 1$^{st}$ name of Akatsuki \citep{Nakamura2007}.}
}
\end{table}

A total of 13 events were confirmed in the set of amateur images sensing the dayside middle clouds of Venus between April and July 2022 (see Table \ref{table:events} and Fig.~\ref{fig:discontinuities}). In average, the discontinuity at the middle clouds exhibited a mean length/width of $4100\pm 800$ / $500\pm 100$ km, while the average phase speed is $c=-84\pm 2$ m$\cdot$s$^{-1}$ and rotation period $5.11\pm 0.09$ terrestrial days. These values match the results from Akatsuki/IR1 images during 2016 ($-74\pm 9$ m$\cdot$s$^{-1}$ and $4.7\pm 0.4$ days; see \citealt{Peralta2020}) and amateur images during March--April 2020 ($-84.9\pm 0.4$ m$\cdot$s$^{-1}$ and $5.08\pm 0.03$ days; see \citealt{Kardasis2022}).\\

The time evolution of the discontinuity zonal phase speed, orientation relative to the equator and meridional coverage is displayed in Figure \ref{fig:CD-properties}. Regarding the night lower clouds, the low number of discontinuity events observed during February 2022 might be related with a potential dissipation of the wave due to the fast increase of the zonal winds after February 4, maybe caused by an episode of equatorial jet \citep{Horinouchi2017NatGeo}. Concerning the events at the dayside middle clouds, two periods may be distinguished. From May to June 2022 the discontinuity exhibits its strongest perturbation over the clouds' albedo, specially between May 30 and June 24 (see Fig.~\ref{fig:discontinuities}) when its characteristic dark-to-bright streak pattern is clearly visible \citep{Kardasis2022}. During this period we also observe that the discontinuity seems to rotate faster as it becomes more perpendicular to the equator (see Fig.\ref{fig:CD-properties}a--b). During July 2022, the rotation period and the orientation exhibit a higher variability, accompanied by a less pronounced alteration on the clouds albedo (see Fig.~\ref{fig:discontinuities}). Although this might be interpreted as an episode of gradual weakening of the discontinuity, we cannot rule out an observational bias since during July 2022 the spatial resolution becomes comparable to the discontinuity width ($500\pm 100$ km).

\begin{figure}
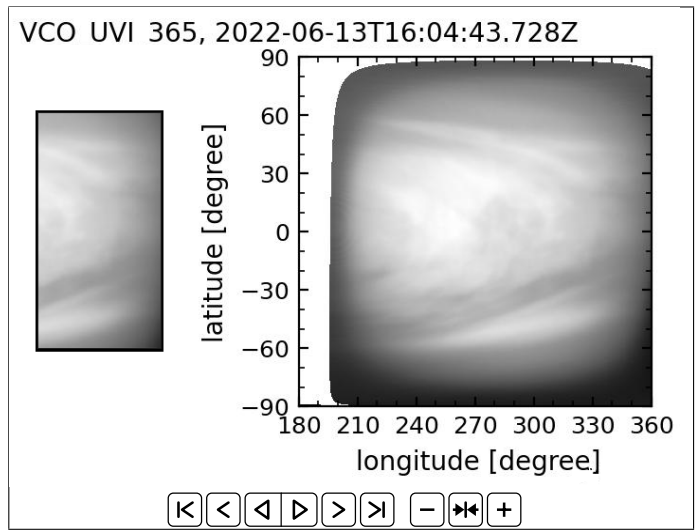

\centering
\begin{frame}{
    \animategraphics[label=disruptions,controls,loop,autoplay,width=0.98\columnwidth]{5}{uvi_CD_evol}{}{}
    \mediabutton[jsaction={if(anim['disruptions'].isPlaying)anim['disruptions'].pause(); else anim['disruptions'].playFwd();}]{}
    }
\end{frame}
\caption{Evolution of the discontinuity-like pattern observed at the clouds' top with Akatsuki/UVI images. This interactive animation exhibits the morphological evolution (left panel) and propagation (right panel) of the discontinuity during the event of its apparent manifestation at the upper and lower clouds (see table \ref{table:events} and figure \ref{fig:discontinuities}). The animation was built with geometrical projections \citep[see level 3 data]{Ogohara2017} of Akatsuki/UVI images taken at 365 nm every about 2 hours from 2022 June 13 at 16:05 UT until June 14 at 23:05 UT. These equirectangular projections consists on longitude–latitude maps with a fixed resolution of 0.125$^{\circ}$ ($2880\times 1440$ grids for 360$^{\circ}$ longitude and 180$^{\circ}$ latitude). The propagation of the discontinuity (right panel) is shown for maps covering latitudes 90$^{\circ}$N--90$^{\circ}$S and longitudes 180$^{\circ}$--360$^{\circ}$. For its morphological evolution (left panel) we used 60$^{\circ}$N--60$^{\circ}$S projections shifted to show the discontinuity at rest. The moving white spot corresponds to dead pixels. No photometric correction was applied.}
\label{fig:uvi-CD-evol}
\end{figure}

\subsection{The discontinuity at the top of the clouds?}\label{propagation-top}
Given that it could not be identified in the images of the upper clouds taken by Akatsuki during 2016, the discontinuity has been interpreted as a Kelvin-type wave which propagates at the deeper clouds and probably dissipates before arriving at the upper clouds \citep{Peralta2020}, which is supported by zonal phase speeds faster than winds at the deeper clouds and slower than winds at the top of the clouds \citep{Kardasis2022}. During June 13--14 (see Table \ref{table:events} and Fig.~\ref{fig:discontinuities}) and probably also in May 30 (see Fig.~\ref{fig:s-CD-up-mid-Clouds}), a discontinuity-like pattern propagating $10\pm 8$ m s$^{-1}$ faster than background winds was observed in UVI images at 365 and 283 nm (Fig.~\ref{fig:discontinuities}), but not in 10-$\mathrm{\mu m}$ thermal images from LIR (presumably due to their worse spatial resolution during these dates). This pattern became apparent at the top of the clouds in June 13 at about 20:00 UT near the border of the --also present-- Y-feature, gradually intensifying and then vanishing in 14 of June 17:00 UT (see animated Fig.~\ref{fig:uvi-CD-evol}).\\
\\
Although the ultraviolet albedo sometimes exhibits sharp patterns, the one on June 13--14 shares with the discontinuity at the middle clouds location, rotation period, orientation and latitudinal coverage, while the zonal winds at the clouds' top are slow enough to change the altitude of the critical level and facilitate the vertical propagation of the Kelvin wave (see Fig.~\ref{fig:CD-properties}). Using the amateur image acquired in 2022 June 14 05:15:50 UT and the UVI image at 06:04:44 UT (see Fig.~\ref{fig:s-CD-up-mid-Clouds}), we derived the position of the discontinuity at the middle clouds when the UVI image was taken considering that $c-84\pm 2$ m$\cdot$s$^{-1}$, and obtained a phase difference $\Delta\varphi =7\pm 2^{\circ}$ (i.e. $\sim 700\pm 200$ km) for the discontinuity at both levels. Assuming that the top of the clouds and the middle clouds are located at $74\pm 1$ km \citep{Ignatiev2009} and $63\pm 5$ km \citep{Khatuntsev2017} respectively, the altitude difference is $\Delta h=11\pm 6$ km and the angle between the wave vector and the horizontal plane ($\tan\phi =\frac{\Delta\varphi}{\Delta h}$) will be $\phi =89\pm 1^{\circ}$. And given that $\tan\phi =\frac{\lambda _{x}}{\lambda _{z}}$ \citep{Gubenko2011} and considering that the width of the discontinuity is representative of the zonal wavelength ($\lambda _{x}=500\pm 100$ km), the vertical wavelength will be $\lambda _{z}=9\pm 2$ km.

\section{Conclusions}\label{conclusions}

   \begin{enumerate}
      \item Thanks to the coordinated observations by JAXA's Akatsuki mission and ground-based professional/amateur observers during the first half of 2022, we present the longest uninterrupted study of the clouds' discontinuity of Venus up to date, extending 109 days from 2022 March 13 until July 31 and covering several cloud layers of the atmosphere.
      \item The nightside lower clouds were explored by means of 2.26-$\mathrm{\mu m}$ images taken by IRTF/SpeX along 41 days from December 2021 until March 2022. Nevertheless, the discontinuity seems to manifest only once prior to an intensification of the zonal winds at lower latitudes.
      \item The middle and upper clouds were studied between 2022 March--July with amateur and Akatsuki/UVI observations, respectively. As many as 13 discontinuity events were identified at the middle clouds, exhibiting an average length and width of $4100\pm 800$ km and $500\pm 100$ km and a mean rotation period of $5.11\pm 0.09$ terrestrial days, a period identical to estimates from 2016 \citep{Peralta2020} and 2020 \citep{Kardasis2022}.
      \item For the first time, we report the possible manifestation of the discontinuity at the top of the clouds during $\sim$21 hours, suggesting that the critical level for this Kelvin wave ascended due to the slower winds at the clouds' top. Nonetheless, further research about atmospheric conditions is required to explain why the discontinuity seemed absent at the clouds' top in other dates with slower winds.
      \item Even though the discontinuity was not identified in Akatsuki/LIR images, future Akatsuki observations might accomplish new positive identifications if discontinuity events occur during pericentric observations (when Akatsuki images have better spatial resolution). In addition, a revisit of Venus ultraviolet images during past episodes of slower winds may reveal new discontinuity events with Akatsuki \citep{Horinouchi2018} and previous space missions like NASA's Pioneer Venus or ESA's Venus Express \citep{Rossow1990,Hueso2015,Khatuntsev2013}, allowing to characterise its recurrence with better accuracy.
   \end{enumerate}

\begin{acknowledgements}
J.P. thanks EMERGIA funding from Junta de Andaluc\'{i}a in Spain (code: EMERGIA20\_00414). I.G.-L. and A.S.-L. were supported by Grant PID2019-109467GB-I00 funded by MCIN/AEI/10.13039/501100011033/ and by Grupos 1128 Gobierno Vasco IT1742-22. We thank the members of JAXA's Akatsuki mission and Visiting Astronomers at the Infrared Telescope Facility operated by the University of Hawaii under contract 80HQTR19D0030 with NASA. This research would have been impossible without the many amateur observers who observed Venus intensively during the investigated period and shared their images, although we would like to highlight contributions from M.~A. Bianchi, G. Calapai, D. Kananovich, N. MacNeill, V. Mirabella, W.~M. Lonsdale, R. Sedrani, L.~S. Viola and G.~Z. Wang. Finally, we thank the two anonymous reviewers whose comments/suggestions helped improve and clarify this manuscript.
\end{acknowledgements}

%
%

\bibliographystyle{aa} 

%

\begin{appendix} 
\section{Discontinuity events projected}
\begin{figure}[h]
\centering
   \includegraphics[width=18cm]{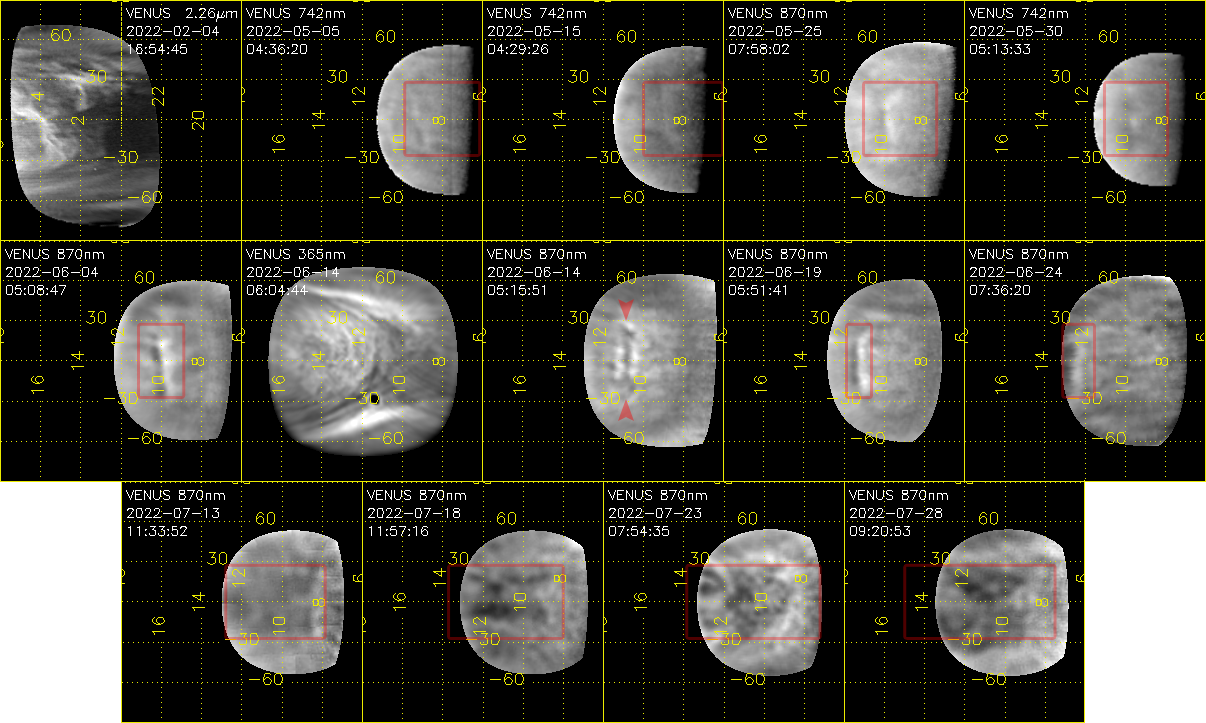}
     \caption{Geometrical projections of discontinuity events during 2022. This Supporting Figure displays the same examples of discontinuity as in Figure \ref{fig:discontinuities} but as equirectanguar (cylindrical) projections with an angular resolution of 0.75$^{\circ}$ per pixel. The images are projected for latitudes 90$^{\circ}$N--90$^{\circ}$S (vertical axis) and local hour coordinates (horizontal axis). The discontinuity was apparent on the nightside lower clouds with $2.26-\mathrm{\mu m}$ images from IRTF/SpeX taken in February 4, and on the dayside middle clouds as observed with images by amateur observers taken with filters covering wavelengths 700--900 nm during May 5,15,25,30, June 14,19,24 and July 13,18,23,28. Exceptionally, the discontinuity seemed to manifest at the top of the dayside clouds with Akatsuki/UVI images at 365 nm during June 14. The red rectangles represent the areas where we expect to find the discontinuity when considering June 14 as the reference position and mean rotation period $5.11\pm 0.09$ terrestrial days.}
     \label{fig:s-discontinuities}
\end{figure}

\clearpage
\section{The discontinuity at the lower clouds}
\begin{figure}[h]
\centering
   \includegraphics[width=\columnwidth]{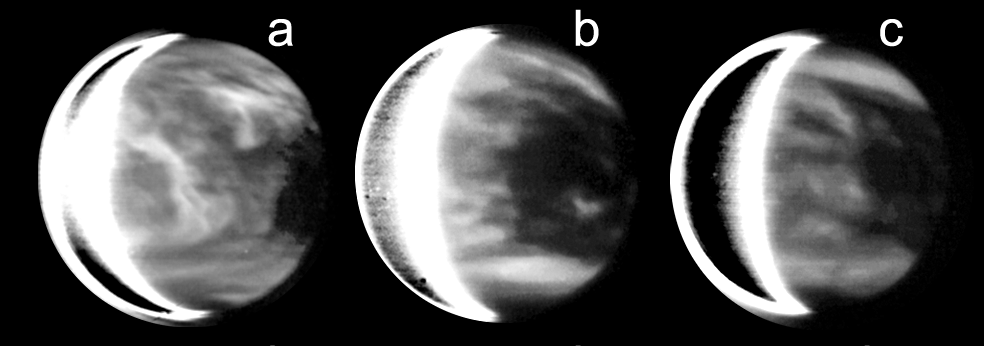}
     \caption{Evolution of the cloud discontinuity at the nightside lower clouds. These $2.26-\mathrm{\mu m}$ images from IRTF/SpeX display the manifestation of the discontinuity at the nightside lower clouds with in 2022 February 4 (a). Even though the abrupt change in the clouds' opacity in February 15 (b) and 20 (c) suggest the reappearance of the first discontinuity after several full revolutions, their zonal speeds match those of the background wind within error bars. The fast increase of zonal wind speeds after February 4 (see Fig.~\ref{fig:CD-properties}a) suggests a probable dissipation of the wave while leaving a persistent reminiscence of its effect on the opacity of the nightside lower clouds.}
     \label{fig:s-CD-LowerClouds}
\end{figure}

\newpage
\section{The discontinuity at two cloud layers}

\begin{figure}[h!]
\centering
\animategraphics[label=cduppermiddle,controls,loop,autoplay,width=0.98\columnwidth]{5}{CD_Upper-Middle_Clouds}{}{}
\mediabutton[jsaction={if(anim['cduppermiddle'].isPlaying)anim['cduppermiddle'].pause(); else anim['cduppermiddle'].playFwd();}]{}
\caption{Clouds' layers during six discontinuity events in 2022. By means of images acquired within less than 1 hour of difference, this interactive animation allows to compare the patterns at the \textit{top of the clouds} (Akatsuki/UVI 365-nm images) when the discontinuity was apparent on images of the \textit{middle clouds}. As in Fig.\ref{fig:s-discontinuities}, 90$^{\circ}$N--90$^{\circ}$S projections in local hour coordinates and resolution of 0.75$^{\circ}$ per pixel are shown. Coordinate grid suppressed for a better visualization.}
\label{fig:s-CD-up-mid-Clouds}
\end{figure}

\end{appendix}

\end{document}